\documentclass[a4paper,12pt]{article}
\usepackage[cp1251]{inputenc}
\usepackage{graphics}
\usepackage{amssymb,amsmath}
\usepackage[english]{babel}
\usepackage{indentfirst}
\begin{document}
\bigskip

\centerline {\bf ON CHROMOSPHERIC VARIATIONS MODELING FOR}
\centerline {\bf MAIN-SEQUENCE STARS OF G AND K SPECTRAL CLASSES}

\bigskip

\centerline {E.A.Bruevich}

\bigskip

\centerline {Sternberg Astronomical Institute, Moscow, Russia}\
\centerline {E-mail: {red-field@yandex.ru}}\

\bigskip
       We present a method of chromospheric flux simulation for 13 late-type main-sequence
       stars. These Sun-like stars have well-determined cyclic flux variations similar
       to $11$ yr solar activity cycle. Our flux prediction is based on chromospheric
       $HK$ emission time series measurements from Mount Wilson
       Observatory and comparable solar data. We show that solar
       three - component modeling explains well the stellar observations. We find that the $10 - 20 \%$
       of K - stars disc surfaces are occupied by bright active regions.

\bigskip


\vskip12pt
{\bf1 INTRODUCTION}
\vskip12pt
            This paper continues the study of variability among Sun-like stars. Here the purpose is
            to obtain the possibility of modeling the behavior of
            the star's chromospheric emission in future or for periods
            of time without measurements.

            Observations of chromospheric variability requires at
            least a decade to reveal variations with timescales
            to the $11$ yr solar cycle.

            We use the data from the observation program that was initiated by Wilson who discovered
            the widespread occurrence of activity cycles by monitoring $CaII$ $H$ and $K$
            variations in 91 stars on or near the lower main
            sequence over $12$ year (Wilson, 1978).
            Two sets of measurements (named "HK-project") have been combined to make
            more than 30 years records of stellar chromospheric
            activity. Wilson made observations from 1966 to 1977
            at monthly intervals on 2.5 m telescope at Mount
            Wilson Observatory. The survey moved in 1977 to 1.5 m
            telescope with instrument whose measurements can be
            compared to those of Wilson's system. Some new stars were added to 91 Wilson's
            stars to bring the total in the monitoring program to 111 stars (Baliunas et al.,
            1995). $Ca II$ $ H $(396.8nm) and $K$
            (393.4nm) emission is observed in
            stars later than approximately $F2 V$, i.e. less massive
            than about $1.5 M_{\odot}$.
                Areas of concentrated magnetic fields on the Sun and
            Sun-like stars emit $Ca II$ $ H $ and $K$ more intensely than areas with less magnetic
            field present. So the contrast of Active Regions (AR)
            emission (where the local magnetic fields are more then some orders higher than
            average global magnetic field) in these $Ca II$ lines
            changes from  $1.2$ to $1.5$ with changing of chromospheric
            activity cycle phase.

            Comparing of variability of $H$ and $K$ emission in
            main-sequence stars should provide important
            validation for theories of magnetic activity, as well
            as place of solar activity in a general perspective.

            The influence of  photospheric flux  in the
            total solar or stars irradiance we can interpretate as the
            cyclic flux variations caused by slight imbalance between
            the flux deficit produced by dark sunspots and the
            excess flux produced by bright faculae.

             Besides of
            such structures as AR in solar and stars chromosphere there is
            another regular structure
            -"chromospheric network" (connecting with the
            supegranulation). It also varies its own
            relative brightness with chromospheric activity cycle.

            We can note that the maximum amplitude of photospheric
            flux variability in $11$ yr solar cycle may be as much
            as $1 - 3\%$ of the average photospheric flux level but the maximum
            amplitude of $Ca II$ chromospheric flux may be as much
            as $20\%$ of the average level. These values are our some estimations for the
            maximum amplitudes of 11 yr variations of Sun-like stars photospheric and chromospheric fluxes.

\bigskip

{\bf2  THE THREE-COMPONENT MODEL OF STELLAR CHROMOSPHERIC EMISSION
AS ANALOG OF TREE-COMPONENT MODEL OF SOLAR CHROMOSPHERIC EMISSION}

\bigskip

            The processes in solar atmosphere caused the emission
            in different spectral intervals and lines are studied
            well enough. But it's very difficult to take into account the contribution of
            all different structures that emitted from the solar
            surface. As a successful example of solar flux model
            calculations in spectral intervals
            of $40 - 140$ nm (that are in agreement with SKYLAB's
            observations) we can point out the (Vernazza et al.,
            1981).

            The Vernazza's calculations take into account
            the influence of 6 main different components on the solar surface
            and their contributions to the total emission in this spectral interval. These
            components are: the dark areas inside the chromospheric network
            cells, the centers of networks, the areas of quiet Sun, the average level of network emission
            region, the bright areas of network, the most bright
            areas of network. When the observations are not made with high
            accuracy all this structures we see as quiet Sun
            average emission (that varies with 11 yr chromospheric cycle).

\begin{figure}[h!]
 \centerline{\includegraphics{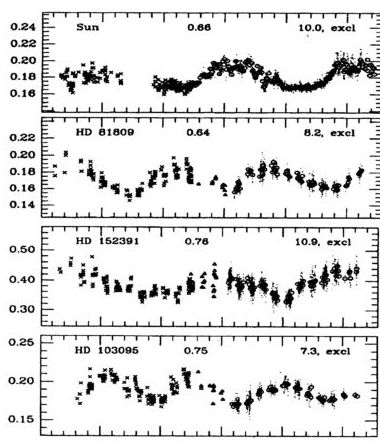}}
 \caption{Records of relative CaII emission fluxes (S-index) from Mount Wilson observations (Baliunas et al.) for stars of "EXELLENT" class for 30 years from 1965 to 1995. HD numbers of stars and their (B-V) values are
presented}\label{Fi:Fig1}
\end{figure}

            These structures contribute significantly
             to the full flux emitted from the quiet Sun
            chromospheric average emission.

            The next most important source of solar
            chromosphere emission is the Active Regions (AR) emission.
            The SKYLAB's observations show the brightness of AR are
            $1.5 - 2.5$ times greater the average quiet surface brightness (Schriver
            et al., 1985). This AR brightness contrast is depend
            of wavelengths. They note also that the AR surface
            brightness depends of the AR area and the number of
            spots the AR consists of.

            Lean's three-component model (that's made for $40 - 140$ nm
            spectral interval) based on NIMBUS 7 observations (Lean et al.,
            1983) assumed that the full flux from chromosphere is determined by three
            main components.

            These components are: (1) - the constant
            component with uniform distributed sources on solar
            surface, (2) - the "active" network component (uniformly
            distributed too but also connected with destroyed parts of previous
            AR and so is proportional to total AR areas), (3) -
            the AR component.

            So one can  use Lean's formulae for calculation of the flux in chromospheric lines:

              $$  I = I_{\lambda Q} \Big\{ 1 + f_N \Big( C_{\lambda N}- 1 \Big) \Big\}
                      + 2 \pi F_ {\lambda Q}(1) \Sigma {A_i \mu_i R_{\lambda}(\mu_i)}
                      \Big ( C_{p \lambda} W_i -1 \Big ) \eqno (1) $$

            where $I$ is the full flux of chromospheric emission, $I_{\lambda Q}$ is the contribution of the
            constant component (BASAL), $C_{p \lambda}$ is the values of AR
            contrasts  and they are similar to contrasts from (Cook et all.,
            1980), $C_{N \lambda}$ is the value of "active network"
            contrast: they are equal to  $0.5 \cdot C_{p \lambda} $
            for continuum and $1/3 \cdot C_{p \lambda}$ for lines,
            $f_{N}$ is part of disk (without AR) that is occupied
            by the "active network".

            The second member in the right part of (1) describes
            emission from all AR on the disk; $A_i$ are values of their squares, $\mu_i$ describes the AR position:
            $\mu_i = {cos {\phi_i} cos {\theta_i}}$ (where $\phi_i$ and
            $\theta_i$ are the coordinates of AR number $i$).
            $R_\lambda(\mu_i)$ describes the relative change of the
            surface brightness $F_{\lambda Q}(\mu_i)$ with moving from
            center to edge of disk. The relative adding AR
            contribution to full flux from the different AR is
            determined by the factor $W_i$ that is linearly changed
            from the value $0.76$ to $1.6$ depending of the
            brightness ball of flocculae (according to ball flocculae changes from $1$ to $5$).

            So the "active" network part in all the surface without AR is determined by the AR decay,
            the next relationship between $f_N$ in time moment $t$
            and average values $A_i$ in earlier time is right:
            $$ f_N(t)= 13.3 \cdot 10^{-5} \cdot < \Sigma A_i(t-27)>  \eqno(2)$$
            where the time-averaging is taken for $7$ previous
            rotation periods, $A_i$ is measured in one million
            parts of the disk.

            To analyzed the $H$ and $K$ $CaII$ flux long-time variations in case of Sun-like stars
             we assume that full flux $S_{CaII}(t)$ is consists of three main components:

              (1) - the"constant part" (so-called BASAL in solar physics - we
            call this component $P_{min}$),

            (2) - the "low-changed phone" (we call this component $P_{CaII}(t)$ ) and

            (3) - "active regions" on the disk of star (we call this component
            $S_{AR}(t)$).

            So the full flux will be $$ S_{CaII}(t) = P_{CaII}(t) + S_{AR}(t)$$
            The component (2) $P_{CaII}(t)$ consists of constant BASAL
            component $P_{min}$ and low-changed pseudo-sinusitis
            component which we can see from the Sun observations
            and will describe it's approximation later.

            It's evident (from solar observations and their interpretations)
             that between the values $S_{CaII}(t)$ and
            $ P_{CaII}(t)$ there is close connection.

            According to (Borovik et al., 1997) the average amplitude of flux
            variations may be $20 \%$ in maximum phase of chromospheric cycle.

            This point of view is according well enough with
            Lean's model (Lean et al., 1983) for solar
            $L_{\alpha}$ line (in case of solar
            $L_{\alpha}$ line flux the maximum amplitude
            of this flux variation in different $11$ yr cycles reached the value of $20 \%$).

            Than we determine the analog coefficient $k$ for star's chromospheric
            cycle as equal to ratio of maximum amplitude of
            so called "phone" component  to maximum amplitude of
            full flux in long-term activity cycle:
            $$k= (P_{CaII}^{max} - P_{min})/(S_{CaII}^{max} -
            P_{min})  \eqno (3)$$

            We consider that $k$ is constant ratio between $ P_{CaII}(t)$
            and $S_{CaII}(t)$ for all moments during star's cycle.

            We also assume that $P_{CaII}^{max}=1.2 \ P_{min}$.

            It's evident from our previous consideration that
            $P_{min}$ is a constant value during all long-term cycles
            but differs for different stars and Sun. Most likely
            that the value $P_{min}$ characterizes  the average
            level of outer atmosphere activity of stars and may
            correlate with ROSAT observations of their X-ray
            fluxes (X-ray luminosity are observed on ROSAT for $65\%$ "HK-project" stars only).

            According to these we connect the full flux value $S_{CaII}(t)$ and
            "phone" flux value $P_{CaII}(t)$ by analog coefficient
            $k$ (3): $$ P_{CaII}(t) = k \cdot S_{CaII}(t) \eqno (4) $$

            The $S_{CaII}(t)$ values we may take from observations
            (Baliunas et al., 1995).

            At Fig 1. we can see
            records of relative $CaII$ $H+K$ emission fluxes ($S_{CaII}$) for 4 from 13 stars
            of EXCELLENT Baliunas's class and Sun for 30 years
            from 1965 to 1995. Also at Fig 1. we can see
            HD numbers of stars and their (B - V) values.

\begin{figure}[h!]
 \centerline{\includegraphics{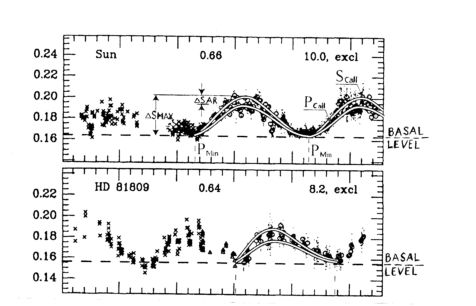}}
 \caption{For Sun and HD 81809 we show BASAL level and other components for (3) - (5) equations}\label{Fi:Fig2}
\end{figure}

            It's evident (from solar observations in different spectral intervals)
            that $P_{CaII}(t)$ and $S_{CaII}(t)$ have
            similar behavior in chromospheric cycle.
            We've calculated the regression coefficients (see Table)
            $a$ and $b$ for the regression relation:
            $$ S_{CaII}(t)= a \cdot P_{CaII}(t) + b \eqno (5) $$

            To make "phone"  flux prediction we use the method
            from (Bocharova et al., 1983) for solar "phone"  flux
            variations in 11 yr cycle. Using Bocharova's
            considerations we've obtained the next approximation
            for "phone"  flux $P_{CaII}(t)$ (Bruevich , 1999):

            $$ P_{CaII}(t)=P_{min}\cdot \Big(1+sin^4 \cdot{ {\pi \cdot t}
            \over{T}} \Big) \cdot e^{-{{\pi \cdot t}
            \over{T}} } \eqno (6) $$

            where $P_{min}$ is the minimum value of "phone" flux
            is equal to BASAL flux. It corresponds to the minimum "phone"
            flux value for the star
            and it's constant for all observed long-term cycles,
            see Fig 2.

            $T$ is the period of long-term chromospheric cycle
            calculated by (Baliunas et al.), $t$ is the time
            expressed in parts of period $T$: $$t=0.1 \cdot
            T, 0.2 \cdot T, ...$$

            So we have two methods of "phone" flux calculations:
            (1) - from observations records using equations (3) and (4) (see
            Fig 2.), (2) - from analytic approximation (6)using $P_{min}$ only. Note that the both methods
            give us very identifiable values of $P_{CaII}(t)$
            which differ some percents only.

            So if we want the chromospheric flux to predict we may calculate
            $S_{CaII}(t)$ with help of equation (5) using $a$ and $b$
            coefficients which are calculated earlier with help of standard
            regression methods and presented in Table.

            In Table we present also the relative
            full flux variation in activity cycle maximum: ($ {\Delta S_{CaII}^{max}}
            / {P_{min}}$) and relative AR adding flux in activity cycle maximum: ($
            \Delta S_{AR}^{max} / P_{min}$).

            These values of flux variations are presented in \% of $P_{mim}$. The value $P_{mim}$ -
            that is equal to BASAL emission for different stars which we can determine from
            Baliunas's data (Fig 1.).
\bigskip

\centerline {Table}

\begin{center}
 \begin{tabular}{|l|c|c|c|c|c|c|c|}
  \hline
  Object & B - V & $T_{cyc}$, yr & $P_{mim}$ & a & b &$ {\Delta S_{CaII}^{max}} / {P_{min}}$& $
            \Delta S_{AR}^{max} / P_{min}$ \\ \hline
  Sun       & 0.66 & 10   & 0.162 & 1.19 & -0.031 & 23.4 & 3.4  \\ \hline
  HD 81809  & 0.64 & 8.2  & 0.155 & 1.13 & -0.020 & 22.6 & 2.6  \\ \hline
  HD 152391 & 0.76 & 10.9 & 0.32  & 1.56 & -0.180 & 31.6 & 11.3 \\ \hline
  HD 103095 & 0.75 & 7.3  & 0.17  & 1.23 & -0.040 & 24.7 & 4.7  \\ \hline
  HD 184144 & 0.80 & 7.0  & 0.19  & 1.45 & -0.085 & 28.9 & 8.9  \\ \hline
  HD 26965  & 0.82 & 10.1 & 0.18  & 1.39 & -0.07  & 27.8 & 7.8  \\ \hline
  HD 10476  & 0.84 & 9.4  & 0.17  & 1.61 & -0.104 & 32.4 & 12.3 \\ \hline
  HD 166620 & 0.87 & 15.8 & 0.175 & 1.43 & -0.075 & 28.6 & 8.6  \\ \hline
  HD 160346 & 0.96 & 7.0  & 0.24  & 1.88 & -0.21  & 37.5 & 17.5 \\ \hline
  HD 4628   & 0.88 & 8.4  & 0.19  & 1.96 & -0.183 & 39.4 & 19.4 \\ \hline
  HD 16160  & 0.98 & 13.2 & 0.19  & 1.61 & -0.116 & 32.6 & 12.6 \\ \hline
  HD 219834B& 0.91 & 10.0 & 0.17  & 1.92 & -0.157 & 38.2 & 18.2 \\ \hline
  HD 201091 & 1.18 & 7.3  & 0.51  & 1.85 & -0.434 & 37.2 & 17.2 \\ \hline
  HD 32147  & 1.06 & 11.1 & 0.22  & 1.67 & -0.147 & 45.4 & 25.4 \\ \hline

 \end{tabular}
\end{center}

            The Table data we may employed in our full flux
            chromospheric predictions: for the certain moment $t$
           ( $t$ is the time expressed in parts of period value $T$)
            we can calculated the value $P_{CaII}(t)$ from
            equation (6). Then from equation (5) for moments $t$ and for $P_{min}$
            star's values we can calculate the predicted flux
            $S_{CaII}(t)$.

\bigskip

{\bf3 SUMMARY AND CONCLUSIONS}

\bigskip

      When we analyze results of our predictions (in Table we presented the observed values
      that we're discussed in this issue and our estimations as $ {\Delta S_{CaII}^{max}}
            / {P_{min}}$ and $\Delta S_{AR}^{max} / P_{min}$) some conclusions can be
      made:

      K-stars of $EXELLENT$ class as Baliunas determined
      for stars with the most evident determination of
      chromospheric activity cycle (Baliunas et al.) have enough number of Active Regions
       at stars surfaces and these AR can emit the addition flux near $10-20\%$ of
      full flux in chromospheric cycle maximum, see Table.
      So we can see that in K-stars the most bright flocculae (its flux is two times
      brighter than the average chromosphere flux) may occupy
      almost $10-20\%$ of star's disk.

      This $10-20\%$ of AR additional flux ("phone" flux evaluation) we show in Fig 2.
      Also we see the star's "phone" flux (smoothly changed in chromospheric
      cycle) and BASAL component (constant in chromospheric cycle).

       Note that all our three components we
      can see in Fig 1. ("HK-project" observations of EXCELLENT stars)
      according to solar case (Lean et al., 1983).

        Acknowledgements. The authors thank the RFBR grant 09-02-01010 for support
         of the work.

\bigskip

{\bf References}

\bigskip

        Baliunas, S.L., Donahue, R.A., et al. (1995). Astrophys.
        J., {\bf 438}, 269.

        Borovik, V.N., Livshitz, M.A., Medar, V.G. (1997).
        Astronomy Reports, {\bf 41}, N6, 836.

        Bruevich, E.A., (1997) Vestnik MSU, Ser3, Physics,Astronomy,
        N6, 48.

        Cook, J.W., Brueckner, G.E., Van Hoosier,
        M.E., (1980) J.Geophys.Res., {\bf A85}, N5, 2257.

        Lean J.L., Scumanich A. (1983) J. Geophys. Res., {\bf
        A88}, N7, 5751.

        Schriver, C.J., Zwaan, C., Maxon, C.W., and Noyes,
        R.W., (1985). Astron. and Astrophys., {\bf 149}, N1, 123.

        Vernazza, J.E., Avrett, E.H., Loeser, R., (1981) Astrophys.
        J.Suppl.Ser., {\bf 45}, N4, 635.

        Wilson, O.C., (1978). Astrophysical J., {\bf 226}, 379.

\end{document}